\documentclass[aps,prc,eqsecnum,showpacs,twocolumn,letterpaper,amssymb,amsmath,nofootinbib]{revtex4-1}
\usepackage{bm}
\usepackage{graphicx}
\usepackage{tabularx}

\begin{document}


\def\EFTNoPions{EFT~$(\hskip+1pt/\hskip-5.7pt\pi)$}
\def\SuperNoPions{\hskip+0.7pt/\hskip-4.5pt\pi\!}
\def\Feynman#1{\hskip+2pt / \hskip-6pt{\bf #1}}
\def\sigmaC{\sigma_{\text{C}}}
\def\sigmaCE1{\sigma_{\text{C}_{E1}}}
\def\sigmaCM1{\sigma_{\text{C}_{M1}}}
\def\tsigmaCM10{\sigma_{\text{C}_{M1(0)}}}
\def\sigmaPD{\sigma_{\text{PD}}}
\def\sigmaPDEs{\sigma_{\text{PD}_{E1}}}
\def\sigmaPDEp{\sigma_{\text{PD}_{E2}}}
\def\sigmaPDMs{\sigma_{\text{PD}_{M1}}}
\def\sigmaTG{\sigma_{2\gamma}}
\def\dibanh{{}}
\def\dibcre{\dagger}
\def\baranh{{}}
\def\barcre{\dagger}
\def\cc#1{\overline{#1}}


%
\title{An empirical model of low-energy $np$ scattering: Insights beyond ERT}


\author{R. W. Hackenburg}
\email[]{hack@bnl.gov}
\affiliation{Physics Department, Brookhaven National Laboratory, Upton, NY 11973} 


\date{\today}

\begin{abstract}
A model is presented of $s$-wave $np$ elastic scattering as proceeding through an intermediate, off-shell dibaryon $d^*$,
with corrections to the $npd$ vertex and $d^*$ propagator.
The model relies on plausible conjectures and hypotheses
to match {\em identically} the form of the simple amplitude given by the shape-independent effective range theory (SI ERT),
which exactly describes the extant $np$ elastic data (within errors) to at least 3~MeV.
The result provides insight into the mechanisms involved in $np$ scattering,
which go beyond what ERT can reveal because ERT is a product of wave-mechanics and is therefore generally independent of mechanism.
For example, in this model,
the signs of the triplet and singlet scattering lengths are determined by the (opposite) spatial symmetries
of the triplet and singlet $np$ wavefunctions and pion exchange in the vertex corrections.
\end{abstract}

\pacs{13.75.Cs,24.50.Dn}

\maketitle

\def\EFTNoPions{EFT~$(\hskip+1pt/\hskip-5.7pt\pi)$}
\def\SuperNoPions{\hskip+0.7pt/\hskip-4.5pt\pi\!}
\def\Feynman#1{\hskip+2pt / \hskip-6pt{\bf #1}}
\def\sigmaC{\sigma_{\text{C}}}
\def\sigmaCE1{\sigma_{\text{C}_{E1}}}
\def\sigmaCM1{\sigma_{\text{C}_{M1}}}
\def\tsigmaCM10{\sigma_{\text{C}_{M1(0)}}}
\def\sigmaPD{\sigma_{\text{PD}}}
\def\sigmaPDEs{\sigma_{\text{PD}_{E1}}}
\def\sigmaPDEp{\sigma_{\text{PD}_{E2}}}
\def\sigmaPDMs{\sigma_{\text{PD}_{M1}}}
\def\sigmaTG{\sigma_{2\gamma}}
\def\dibanh{{}}
\def\dibcre{\dagger}
\def\baranh{{}}
\def\barcre{\dagger}
\def\cc#1{\overline{#1}}

\section{Introduction}
\label{sec1}

\begin{figure}
\includegraphics[width=8.6cm]{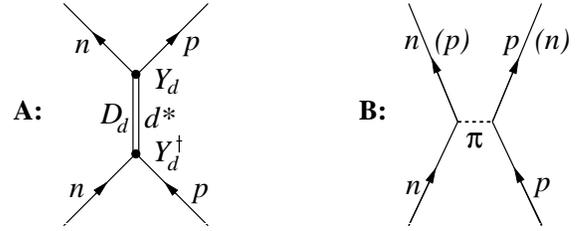}
  \caption{
  \label{Fig1}
  {\textbf A}: 
  Uncorrected effective-field amplitude: $d^*$ propagator $D_d$, $npd$ vertex operator $Y_d$.
  {\textbf B}: Single pion exchange.}
\end{figure}

Shortly after the discovery of the neutron, Wigner
determined the $np$ elastic
scattering (partial) cross section essentially as $1/m_N(T+\epsilon_d)$ times an effective range correction,
where $m_N$ is the nucleon mass,
$T$ is the $np$ c.m. kinetic energy, and where $\epsilon_d$ is the binding energy
either of the deuteron
or
of a virtual singlet $np$ state\cite{754,138}.
This result is very nearly identical to that of
the shape-independent
effective range theory (SI ERT),
which came into use in the late 1940s\cite{365,366} and is still in use today.
At low energy, the factor $1/m_N(T+\epsilon_d) \cong 4/(E^2-m_d^2)$, where $E$ is the total $np$ relativistic energy and
$m_d = m_n+m_p - \epsilon_d$.
This is suggestive of a propagator of pole mass $m_d$.
In effective-field theoretic treatments employing dibaryons (dEFT),
(e.g., Refs. \cite{1838,1859,1841,1896,1861,2141}),
low-energy neutron-proton elastic scattering is treated
as if it proceeded through an intermediate dibaryon $d^*$,
as shown in Fig. \ref{Fig1}A.
There are two ($np$) dibaryons: the spin-triplet ($J=1$) and the spin-singlet ($J=0$),
with masses $m_t$ and $m_s$ corresponding to the two shallow $s$-wave $np$ scattering poles.

A diagrammatic model of low-energy $np$ scattering is presented
which yields an amplitude that matches identically the form of the SI ERT amplitude
as expressed in terms of the scattering length and effective range.
This matching is accomplished with the aid of a few hypotheses and conjectures
which are based on plausible physical mechanisms and arguments.
The result permits the uncorrected propagator and, separately,
the corrections to the $npd$ vertex and $d^*$ propagator to be lifted directly from SI ERT,
thereby enabling several observations to be made regarding mechanism.
Natural units ($\hbar=c=1$) are used throughout.

\section{The Constructed Amplitude}
\label{sec2}
Let $E$ be the total c.m. energy of a two-particle scattering system.
Let $P_i = p_1+p_2$ and $P_f = p_3+p_4$,
where $p_1, p_2$ are the 4-momenta of the initial particles and $p_3, p_4$ those
of the final particles.
In the c.m.,
$|{\bm p}_1| = |{\bm p}_2| \equiv p_i$, $|{\bm p}_3| = |{\bm p}_4| \equiv p_f$,
and $P_i = P_f = (-E,0)$.
With initial relative velocity
$v = p_i E/p_1^0 p_2^0$,
the normalized, covariant flux-and-phase-space factor is \cite{1232}
\begin{gather}
\label{2.1}
\frac{(2\pi)^2}{2p_1^0 2p_2^0 v} \! \int \!\! \frac{d^3{\bf p}_3}{2p_3^0}\frac{d^3{\bf p}_4}{2p_4^0}
\, \delta^{(4)}(P_f\!-\!P_i)
= \frac{\pi^2}{4E^2} \frac{p_f}{p_i} \, d\Omega \ .
\end{gather}
For elastic scattering, $p_i = p_f \equiv p$.
Let $\left|np_i \right>$ and $\left|np_f \right>$ be
initial and final $np$ non-interacting two-particle wavepacket states.
For a matrix element $\left<np_f \right| {\cal T} \left|np_i \right>$,
the $s$-wave elastic differential cross section is
\begin{equation}
\label{2.2}
d\sigma/d\Omega = (\pi^2/4 E^2) \left| \left<np_f \right| {\cal T} \left|np_i \right> \right|^2 \ .
\end{equation}

In what follows, $d$ represents either $t$ or $s$
for the spin-triplet or spin-singlet off-shell dibaryon.
Let $Y_d$ (shorthand for $Y_{npd}$) be the $npd$ vertex operator,
where the lone subscript indicates the off-shell particle.
Let $D_d$ be the dibaryon propagator, and 
let $\left| d^* \right> \left< d^* \right|$ be a projection onto the dibaryon state.
A convenient notation combines $D_d$
with $\left| d^* \right> \left< d^* \right|$ to form a propagator-projection, thus,
\begin{equation}
\label{2.3}
{\bm D}_d \equiv \left| \smash{d^*} \right> D_d \left< \smash{d^*} \right| \ .
\end{equation}
With vertex correction $v_d$ (real, as will be shown) and propagator correction $q_d$ (complex), both scalars,
the constructed amplitude is
\begin{equation}
\label{2.4}
A_d=\left<np_f \right| {\cal T} \left|np_i \right>
=\left<np_f \right| v_d Y_{d}^\dibanh {\bm D_d} q_d Y_{d}^\dibcre v_d \left|np_i \right> \ .
\end{equation}
This {\em unitless\/} amplitude does not include the flux-and-phase-space factor,
which would appear as the square root of the factor $\pi^2/4 E^2$ from Eq.~\eqref{2.2}.
Separating this factor from the amplitude simplifies the extension to related inelastic channels,
where this factor is, more generally, $(\pi^2/4E^2) p_f/p_i$.

With Eq.~\eqref{2.3}, the amplitude Eq.~\eqref{2.4} is
\begin{equation}
\nonumber
A_d = v_d^2 q_d \left<np_f \right|Y_d^\dibanh \left|\smash{d^*}\right> D_d
\left< \smash{d^*} \right| Y_d^\dibcre \left|np_i \right> \ .
\end{equation}
$Y_d^\dibanh$ annihilates a $d^*$, creates an $np$, and is characterized by the
single-vertex transition element $y_d$, thus,
\begin{gather}
\label{2.5}
\left< np \right| Y_d^\dibanh \left| \smash{d^*} \right> = y_d \ , \ \ 
\left< \smash{d^*} \right| Y_d^\dibcre \left| np \right> = \cc{y}_d \ .
\end{gather}

The quantity $|y_d|^2$ is a transition rate, and has units of energy.
But what energy?
The simplest (and perhaps obvious) choice is the energy of the off-shell leg, which is the energy
of the mediating field, i.e., the dibaryon, and which
also happens to be the total energy of the $np$ system.
Given that $Y_d$ has no angular dependence,
$y_d$ contains the $s$-wave spherical harmonic $Y_{0}^{0}=1/\sqrt{4\pi}$.
By hypothesis, define\footnote{
  With Eq.~\eqref{2.5} as a normalized overlap integral, this corresponds to 100\% overlap between the
$\left| \smash{d^*} \right>$ and $\left| np \right>$ wavefunctions.
}
\begin{gather}
\label{2.6}
y_d \equiv Y_{0}^{0} \sqrt{E} = \sqrt{E/4\pi} \ .
\end{gather}
The amplitude is then
\begin{equation}
\label{2.7}
A_d = v_d^2 q_d y_d D_d \cc{y}_d = v_d^2 q_d D_d E/4\pi \ .
\end{equation}

\section{The Vertex Correction}
\label{sec3}
\begin{figure}
\includegraphics[width=8.6cm]{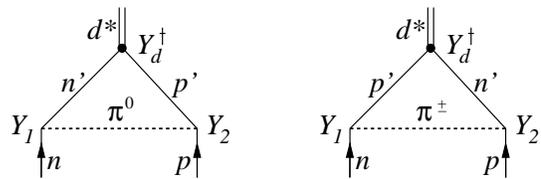}
  \caption{
  \label{Fig2}
  The first-order corrections at the initial $npd$ vertex due to $\pi^0$ and $\pi^\pm$ exchange.
  The corrections at the final vertex are the same.
  $Y_1$ and $Y_2$ are $NN\pi$ vertex operators.
  }
\end{figure}

Consider an infinite sequence of diagrams formed by 
Fig. \ref{Fig1}A joined to chains of increasing length, formed of Fig. \ref{Fig1}B.
The first-order terms in this sequence are shown in Fig. \ref{Fig2}.
The energy and total momentum of the $np$ must be the same at the $npd$ vertex as in the initial or final non-interacting state,
and this is true for any number of pion exchanges between the $np$ at a vertex.
These pion exchanges therefore have no effect on $Y_d$.
The effect of pion exchanges between the $np$ legs is simply to add
diagrams to the amplitude, which will either increase or decrease the amplitude, depending
on the phase of the contributions from these extra diagrams.
All that is needed is the probability that a single exchange will occur,
along with the relative phase of the $np$ after the exchange.
The pion-nucleon interaction is spin-independent,
so an exchange comprising a particular pion and its corresponding pair of $NN\pi$ transition elements 
contributes to the amplitude with exactly the same magnitude
whether the $np$ is in the triplet or singlet spin state
(though not necessarily with the same phase).
\begin{itemize}
\item[]
{\bf Conjecture I}: Pion-exchange between the $np$ legs near an $npd$ vertex
  may be implemented with a pair of commuting operators $H_0$ and $H_1$ acting on $|np\big>$,
  with scalar eigenvalues $h_0$ and $h_1$,
  where $H_0$ corresponds to $\pi^0$ exchange and $H_1$ to $\pi^\pm$ exchange.
  These eigenvalues (squared) correspond to the probability that a $\pi^0$ or $\pi^\pm$ exchange occurs,
  and their phases give the relative change in the phase of the $np$ wavefunction.
\end{itemize}
\begin{figure}
\includegraphics[width=8.6cm]{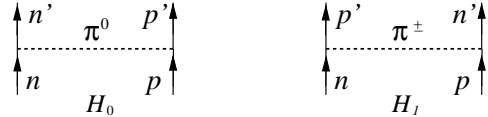}
  \caption{
  \label{Fig3}
  The pion exchange operators $H_0$ and $H_1$.
  $H_1$ inverts the phase of the singlet $\left|np \right>$, where the spatial wavefunction is antisymmetric,
  but preserves the phase of the triplet.
  $H_0$ always preserves the phase because there is no $np$ exchange.
  }
\end{figure}
If $h_0$ and $h_1$ are obtained empirically, as is done here, then
they will include other (e.g., non-pionic) exchanges,
but at low energies it is reasonable to ignore non-pionic exchanges.
When necessary to distinguish between $\pi^-$ and $\pi^+$ exchange,
take $H_1 = H_- + H_+$ and $h_1 = h_- + h_+$,
but for what follows this would add unnecessary clutter.
The important distinction is between $H_0$ and $H_1$ (Fig. \ref{Fig3}),
because neutron-proton exchange occurs under $H_1$, but not under $H_0$, thus,
\begin{align}
\label{3.1}
H_0 |np\big> &= h_0|np\big> \qquad \mbox{(no $np$ exchange)},
\\
\label{3.2}
H_1|np\big> &=
\begin{cases}
+h_1|np\big> & \mbox{triplet} \\
-h_1|np\big> & \mbox{singlet}
\end{cases} \ (np\  \mbox{ exchange}),
\end{align}
where the signs in the $np$ exchange follow the spatial symmetries of the
triplet (symmetric) and singlet (antisymmetric).
By making the factor $\pm 1$ from $np$ exchange explicit in Eq.~\eqref{3.2},
the same eigenvalue $h_1$ can be used for both singlet and triplet.
The series containing all orders (including zero) of pion exchange at one vertex is
\begin{align}
\nonumber
&\phantom{=} (1+H_0+H_1+H_0^2+2H_0H_1+H_1^2 + ...) |np\big>
\\ \nonumber
&= (1+(h_0 \pm h_1)+(h_0 \pm h_1)^2+(h_0 \pm h_1)^3 ...) |np\big>
\\ \label{3.3}
&= \begin{cases}
(1-h_0-h_1)^{-1}\ |np\big> & \ \rm{triplet} \\
(1-h_0+h_1)^{-1}\ |np\big> & \ \rm{singlet}
\end{cases} \ ,
\end{align}
provided that $|h_0 \pm h_1| < 1$.
The vertex corrections are
\begin{equation}
\label{3.4}
v_t \equiv (1-h_0-h_1)^{-1} \ , \ \  v_s \equiv (1-h_0+h_1)^{-1} \ .
\end{equation}

%
\section{The propagator correction}
\label{sec4}
\begin{figure}
\includegraphics[width=8.6cm]{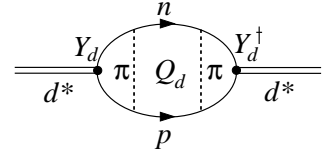}
  \caption{
  \label{Fig4}
  The first-order $d^*$ propagator correction.
  It includes the uncorrected propagator of {\em one\/} of the two $d^*$ shown.
  $Q_d$ is a double-nucleon propagator, and includes the pion exchange correction.
  The two pions shown represent
  an infinite sequence of pion exchanges.
  }
\end{figure}

The first-order propagator correction is shown in Fig. \ref{Fig4}.
Because $E>m_n+m_p$, the neutron and proton in the loop are on-shell.
Let the correction due to pion exchange in each loop of the propagator correction be given by $V_d$,
which is different than $v_d\,$ (or $v_d^2$):
Whereas pion exchange need not always occur at the initial and final (external) $npd$ vertices,
it is {\em obligatory\/} in the loop,
because if there is no pion exchange between the $np$ in the loop,
then they will escape.
A pion exchange does not guarantee that the $np$ loop will close, but
without any exchanges, they certainly will not close.
By hypothesis, each pion exchanged in the loop is associated with an internal $npd$ vertex, which may be either vertex.
Because the $np$ are on-shell, the same eigenvalues $h_0$ and $\pm h_1$ characterizing pion exchange at an
external $npd$ vertex should apply equally well to pion exchange at an internal $npd$ vertex.
Let $L \equiv h_0 \pm h_1$ be the first-order pion-exchange contribution from the ``left'' vertex, and 
$R \equiv h_0 \pm h_1$ that from the ``right'' vertex (upper sign for the triplet).
Then the lowest order contribution is $L + R = 2(h_0 \pm h_1)$,
the next is $L^2 + L R + R^2 = 3(h_0 \pm h_1)^2$, etc.
Instead of Eq.~\eqref{3.3}, the expansion in the loop is
\begin{equation}
\nonumber
(L\!+\!R) + (L^2\!+\!L R\!+\!R^2)
+ (L^3\!+\!L^2\!R\!+\!L R^2\!+\!R^3) + ... \ ,
\end{equation}
which yields for $V_d$ the series (upper sign for the triplet)
\begin{align}
\nonumber
V_d
&= \sum_{n=1}^\infty (n+1)(h_0 \pm h_1)^n
\\ \nonumber
&= \frac{d}{d(h_0\pm h_1)}\left[\sum_{n=0}^\infty (h_0 \pm h_1)^n - (h_0 \pm h_1) - 1 \right]
\\ \nonumber
&= \frac{d(1 - h_0 \mp h_1)^{-1}}{d(h_0\pm h_1)} - 1
= (1 - h_0 \mp h_1)^{-2} - 1
\\ \label{4.1}
&= v_d^2 - 1 \ .
\end{align}
This is just the vertex correction Eq.~\eqref{3.4} squared,
with the uncorrected contribution (unity) subtracted off.
\begin{itemize}
\item[]
{\bf Conjecture II}:
  The propagator correction consists of a series of frustrated attempts
  by the $np$ to break away and return to the free state,
  but forced to close by pion exchanges in the loop.
  The correction therefore contains the phase space $\int_{4\pi} d\Omega (p/4E) = \pi p/E$,
  applied to the opening vertex of each loop.
  Because there is no loop if the $np$ do not close, there is no need for
  a flux factor at the closing vertex.
  Including the effect of the pion exchanges,
  the contribution from each loop contains the factor $(v_d^2-1)(\pi p/E)$.
\end{itemize}
By hypothesis, define the double-nucleon propagator-projection, including the pion-exchange correction, to be
\begin{equation}
\label{4.2}
{\bm Q}_d \equiv \left| n p \right> Q_d \left< n p \right| \qquad
Q_d \equiv (v_d^2-1) (\pi p/E) \ .
\end{equation}
Inserting the entire series of corrections into Eq.~\eqref{2.4} in place of $q_d$,
\begin{align}
\nonumber
A_d &= \left<np_f \right| v_d Y_d^\dibanh
\big( {\bm D}_d
+ {\bm D}_d Y_d^\dibcre {\bm Q}_d Y_d^\dibanh {\bm D}_d +
\\ \nonumber
& + {\bm D}_d Y_d^\dibcre {\bm Q}_d Y_d^\dibanh {\bm D}_d \ 
  Y_d^\dibcre {\bm Q}_d Y_d^\dibanh {\bm D}_d + ... \ \big)
Y_d^\dibcre v_d \left|np_i \right> \ .
\end{align}
Arbitrarily designating the propagator ${\bm D}_d$ on the left of each term as the uncorrected propagator
and factoring it out on the left,
and allowing for the possibility that the dibaryon propagators in the correction
might be different than the uncorrected propagator,
with ${\bm D}_{d0} \equiv \left| d^* \right> D_{d0} \left< d^* \right|$ the dibaryon propagator-projection in the correction,
yields
(the notation anticipates a result from Sec. \ref{sec6})
\begin{align}
\nonumber
A_d &= \left<np_f \right| v_d Y_d^\dibanh {\bm D}_d
\big( 1 +
Y_d^\dibcre {\bm Q}_d Y_d^\dibanh {\bm D}_{d0} +
\\ \nonumber
&\phantom{=} + \! Y_d^\dibcre {\bm Q}_d Y_d^\dibanh {\bm D}_{d0} \ 
  Y_d^\dibcre {\bm Q}_d Y_d^\dibanh {\bm D}_{d0} + ... \big)
Y_d^\dibcre v_d \left|np_i \right>
\\ \nonumber
 &= v_d^2 \left<np_f \right| Y_d^\dibanh
\left| d^* \right> D_d
\big( \left< d^* \right| +
\\ \nonumber
&\phantom{=}+\left< d^* \right| Y_d^\dibcre \left| np \right>
Q_d \left< np \right| Y_d^\dibanh \left| d^* \right>
D_{d0} \left< d^* \right|
+ ...
\big)
Y_d^\dibcre \left|np_i \right>
\\ \nonumber
 &= v_d^2 |y_d|^2 D_d \big( 1 + |y_d|^2 Q_d D_{d0} + |y_d|^4 Q_d^2 D_{d0}^2 + ...  \big)
\\ \label{4.3}
&= v_d^2 (E/4\pi) D_d \left[ 1 - p (v_d^2-1) D_{d0}/4 \right]^{-1} \ ,
\end{align}
where the last step uses Eqs.~\eqref{2.6}, \eqref{4.2}, and requires
\begin{equation}
\label{4.4}
|p (v_d^2-1) D_{d0}/4|<1 \, .
\end{equation}
%
Define the (scalar) propagator correction to be
\begin{align}
\label{4.5}
q_d &\equiv [ 1 - p (v_d^2-1) D_{d0} /4 ]^{-1} \ .
\end{align}

A contribution from the $d$-wave for the triplet
can be ignored at $T \lesssim 3$~MeV,
or even $T\lesssim 10$~MeV to a very good approximation \cite{2000,217}.

\section{Singlet-Triplet Interference}
\label{sec5}

For unpolarized beams and targets, there are three equal contributions to the amplitude
from the triplet, and one from the singlet;
ordinarily, none of these interfere.
For very low energies, the scattering may be coherent.
In this case, for nearby pairs of scatterers in opposite spin-states,
the four $\left|{J,M}\right>$ states are equally populated and interfere.
The coherent amplitude is \cite{837}
\begin{equation}
\label{5.1}
2 \left( \tfrac{3}{4} A_t + \tfrac{1}{4} A_s \right)
= \left( \tfrac{3}{2} v_t^2 q_t D_t + \tfrac{1}{2}v_s^2 q_s D_s \right) E/4\pi \ .
\end{equation}

For polarized beams and targets, the $M=\pm 1$ amplitudes are pure triplet,
but the $M=0$ amplitude is an interfering mixture of singlet and triplet, thus,
\begin{equation}
\label{5.2}
A_t + A_s
= \left( v_t^2 q_t D_t + v_s^2 q_s D_s \right) E/4\pi \ .
\end{equation}

Under the phase normalization convention of Ref. \cite{1058}
for the construction of a two-particle helicity state
from two single-particle helicity states for spin-$\tfrac{1}{2}$ particles,
the $np$ wavepacket-state phase normalization is
$(-1)^{\frac{1}{2}-\frac{1}{2}}=1$ for the triplet and
$(-1)^{\frac{1}{2}-(-\frac{1}{2})}=-1$ for the singlet, or, equivalently,
$(-1)^{J-1}$ for both.
These phase-normalizations might seem to cancel in the matrix element $\left<np_f \right|{\cal T} \left|np_i \right>$,
but if the phase of the intermediate dibaryon is to be continuous with the initial and final free $np$ states,
then the dibaryon propagator must also have this same phase-normalization, thus,
\begin{equation}
\label{5.3}
D_d \propto (-1)^{J-1} \ .
\end{equation}
The well-known destructive interference between the
triplet and singlet amplitudes in Eqs.~\eqref{5.1} and \eqref{5.2}
then follows from their opposite phase normalizations,
which in turn follows from the spatial symmetries of the free $s$-wave $np$ wavefunctions.

\section{Comparison with ERT}
\label{sec6}

The ERT partial amplitude (which implicitly includes the phase space) is
\begin{equation}
\label{6.1}
f_d = -1 \big/ \left(i p + a_d^{-1} - \tfrac{1}{2}\,r_d\,p^2 \right) \ ,
\end{equation}
where $p$ is the $np$ c.m. momentum,
$a_d$ and $r_d$ are the scattering length and effective range,
and where $d$ is either $t$ or $s$ for the spin-triplet or spin-singlet.
There are two poles (each, for triplet and singlet) in the amplitude Eq.~\eqref{6.1}.
Both lie on the imaginary axis of the complex $p$-plane,
at $p = i\gamma_d$ and $p = i\beta_d$, given by
\begin{align}
\label{6.2}
\gamma_d = r_d^{-1} \left(1 - \sqrt{1-2r_d/a_d}\right)\ ,
\\ \label{6.3}
 \beta_d = r_d^{-1} \left(1 + \sqrt{1-2r_d/a_d}\right)\ .
\end{align}
These solutions are assigned so that $|\gamma_d|^2 < |\beta_d|^2$.
Because $2r_d < |a_d|$, both $\gamma_d$ and $\beta_d$ are real if $a_d$ is real.

In terms of the poles, Eq.~\eqref{6.1} is
\begin{equation}
\label{6.4}
f_d = \frac{ - \gamma_d \beta_d a_d }{( \gamma_d + i p )( \beta_d + i p)}
= \frac{ - \gamma_d a_d }{( \gamma_d + i p )(1+ i p/\beta_d)} \ .
\end{equation}
Define the {\em effective range correction\/} to be
\begin{equation}
\label{6.5}
R_d \equiv \gamma_d a_d \big/ \left(1 \! + \! i p/\beta_d \right) \ .
\end{equation}
Then, for real $p$ and $\varphi_d$,
\begin{equation}
\label{6.6}
f_d = - R_d e^{i \varphi_d} \big/ \sqrt{p^2 + \gamma_d^2} \ .
\end{equation}
The $np$ c.m. kinetic energy is $T \equiv E - m_n - m_p$.
At the pole $p = i \gamma_d$, we have
$p^2 = -\gamma_d^2 < 0$, which corresponds to $T=-\epsilon_d<0$ \cite{2000,1551} and $E = m_d \equiv m_n+m_p-\epsilon_d$,
where $\epsilon_d$ is the $np$ binding energy and $m_d$ is the pole mass.
(Nonrelativistically, $\gamma_d^2 = 2m_{np}\,\epsilon_d$, with reduced $np$ mass $m_{np}$.)
The relativistic, kinematic relation between $E$ and $p^2$ relates
$m_d$ to $\gamma_d^2$, thus,
\begin{align}
\nonumber
p^2 &= \tfrac{1}{4}[E^2 - 2(m_n^2+m_p^2) + (m_n^2 - m_p^2)^2/E^2] \ ,
\\ \label{6.7}
-\gamma_d^2 &= \tfrac{1}{4}[m_d^2 - 2(m_n^2+m_p^2) + (m_n^2 - m_p^2)^2/m_d^2] \, .
\end{align}
With $n_d \equiv 1-(m_n^2-m_p^2)^2/E^2 m_d^2$ and Eq.~\eqref{6.7},
\begin{equation}
\label{6.8}
p^2+\gamma_d^2=\tfrac{1}{4} n_d (E^2-m_d^2) \ .
\end{equation}
Since $n_d$ differs from unity by less than $5\times 10^{-7}$ ($n_d \to 1$ as $E \to \infty$), it will be dropped.
Then
\begin{equation}
\label{6.9}
f_d = - 2 R_d e^{ i \varphi_d} \big/ \sqrt{\smash[b]{E^2-m_d^2}} \ .
\end{equation}

Equating $f_d$ to $A_d$ times the square root of the phase space factor in Eq.~\eqref{2.2}
yields $(\pi/2E) A_d = f_d$.
Then, from Eqs.~\eqref{6.9} and \eqref{2.7},
\begin{equation}
\label{6.10}
 \tfrac{1}{8} v_d^2 q_d D_d = - 2 R_d e^{i\varphi_d} \big/ \sqrt{\smash[b]{E^2-m_d^2}} \ .
\end{equation}
By hypothesis, the effective range correction $R_d$ is identified with the vertex and propagator corrections.
Then
\begin{align}
\label{6.11}
v_d^2 q_d &= R_d \ ,
\\ \label{6.12}
\tfrac{1}{8} D_d &= -2 e^{i\varphi_d} \big/ \sqrt{\smash[b]{E^2-m_d^2}} \ .
\end{align}
Adopting a phase-normalization convention
to agree with Eq.~\eqref{5.3}, i.e.,
$-e^{i\varphi_d} = (-1)^{J-1} e^{i\varphi}$, with $\varphi$ an undetermined phase
(the {\em same\/} for both triplet and singlet),
\begin{align}
\label{6.13}
D_d &= (-1)^{J-1} 16 \, e^{i\varphi} \big/ \sqrt{\smash[b]{E^2 \!- m_d^2}}
\\ \label{6.14}
    &= (-1)^{J-1} 16 \, e^{i\varphi} \big/ \sqrt{\smash[b]{E_{d^*}^2 \!- p_{d^*}^2 \!-m_d^2}} \ ,
\end{align}
where $p_{d^*}=0$ and $E_{d^*}=E$ in the rest frame of the $d^*$.

Equation~\eqref{6.14} is a curious form for a propagator
-- usually, one sees this form {\em without\/} the radical.
This may be attributed to the timelike nature of the intermediate state,
which should actually have a finite lifetime consistent with the uncertainty principle.
I.e., for some brief period $\sim 1/2(E-m_d)$, $E$ is indistinguishable from $m_d$,
so the propagator applies to a particle (amplitude-squared),
and not simply the amplitude.
To correctly incorporate such a propagator into the amplitude, it is necessary to take
the square root, in the same fashion as the square root of the phase space must be
employed when it is incorporated into an amplitude.
The factor 16 suggests an unidentified counting rule,
though it could also, at least in part, reflect the choice of normalization in Eq.~\eqref{2.1}.

From Eqs.~\eqref{6.5} and \eqref{6.11}, the zero-energy correction is, with $q_d(p=0)=1$,
\begin{equation}
\label{6.15}
v_d^2 = R_d(p=0) = \gamma_d a_d \ .
\end{equation}
From Eqs.~\eqref{6.2}, \eqref{6.3}, and \eqref{6.15},
\begin{equation}
\label{6.16}
\beta_d = \gamma_d/(\gamma_d a_d - 1) = \gamma_d/\left(v_d^2 - 1 \right) \ .
\end{equation}
From Eqs.~\eqref{6.5}, \eqref{6.11}, \eqref{6.15}, and \eqref{6.16},
\begin{equation}
\label{6.17}
q_d = 1/(1+ i p/\beta_d)
 = [ 1+ i p\left(v_d^2 - 1 \right)/ \gamma_d ]^{-1} \ .
\end{equation}
The two poles at $i\beta_d$ given by Eq.~\eqref{6.3},
which have been called ``physically meaningless'' \cite{1925} or ``unphysical deep'' \cite{1845} poles,
have a physical interpretation in this model:
They are momentum scales in the propagator correction,
determined by the pole masses (through $\gamma_d$)
and the obligatory pion exchange ($v_d^2 - 1$) in the propagator correction.

Comparing Eqs.~\eqref{4.5} and \eqref{6.17} reveals
\begin{equation}
\label{6.18}
D_{d0} = - 4 i / \gamma_d \ .
\end{equation}
From Eq.~\eqref{6.8}, where $E=m_n+m_p$ for $p=0$, and dropping $n_d$,
\begin{equation}
\label{6.19}
\gamma_d^2 = \tfrac{1}{4} \left[ (m_n+m_p)^2-m_d^2 \right] \ .
\end{equation}
Because the sign of $D_{d0}$ is unimportant in Eq.~\eqref{4.5}, the
only requirement being that $D_{d0}$ is pure imaginary,
the phase convention from Eq.~\eqref{6.14} is adopted, yielding
\begin{align}
\label{6.20}
D_{d0} &= (-1)^{J-1} \, 8 \, e^{i\varphi} \big/\sqrt{\smash[b]{(m_n+m_p)^2 - m_d^2}} \ ,
\end{align}
which requires $\varphi=\pm\tfrac{\pi}{2}$.
Thus, $D_{d0}$ is simply $\tfrac{1}{2} D_d$ at $p=0$.
Why $D_{d0}$ would be independent of the energy is something of a puzzle,
but this is how the form of SI ERT determines it to be.
The factor $\tfrac{1}{2}$ (i.e., 8 vs. 16) suggests an unidentified counting rule.
Using Eqs.~\eqref{6.18} and \eqref{6.20} to determine the relation between $e^{i\varphi}$
and the sign of $\gamma_d$ as given by Eq.~\eqref{6.19} gives
\begin{align}
\label{6.21}
\gamma_d &= -i \tfrac{1}{2} (-1)^{J-1} \, e^{-i\varphi} \sqrt{\smash[b]{(m_n+m_p)^2-m_d^2}} \ ,
\end{align}
which is consistent with Eq.~\eqref{6.2}, i.e. $\gamma_d>0$ for $J=1$ (triplet) and $\gamma_d<0$ for $J=0$ (singlet),
if and only if $\varphi=- \tfrac{\pi}{2}$,
which hereafter will be adopted.
Then,
\begin{align}
\label{6.22}
D_d &= i (-1)^J \, 16 \big/ \sqrt{\smash[b]{E_{d^*}^2 \!- p_{d^*}^2 \!-m_d^2}} \ ,
\\ \label{6.23}
D_{d0} &= i (-1)^J \, 8 \big/\sqrt{\smash[b]{(m_n+m_p)^2 - m_d^2}} \ .
\end{align}

The SI ERT parameters from Ref. \cite{2000} are
\begin{align}
\nonumber
& a_t = \phantom{-2} 5.4112(15)~\mbox{fm} \quad & r_t = 1.7436(19)~\mbox{fm}
\\ \label{6.24}
& a_s = -23.7148(43)~\mbox{fm} \quad & r_s = 2.750(59)~\mbox{fm} \ .
\end{align}
With Eqs.~\eqref{6.2}, \eqref{6.15}, and \eqref{6.24},
the vertex corrections are
\begin{equation}
\label{6.25}
v_t = 1.119~33(16) \ , \qquad
v_s = 0.973~60(53) \ .
\end{equation}
The non-exchange and exchange eigenvalues 
$h_0$ and $h_1$ can be solved for in terms of $v_t$ and $v_s$ from Eq.~\eqref{3.4}, thus,
\begin{align}
\label{6.26}
h_0 & = 1 - \tfrac{1}{2}\left( v_s^{-1} + v_t^{-1} \right)\!\!\!\!\!\!\!\!\!\!\!&=& \, 0.039~75(28) \ ,
\\ \label{6.27}
h_1 & = \tfrac{1}{2}\left( v_s^{-1} - v_t^{-1} \right)                          &=& \, 0.066~86(28) \ .
\end{align}
These satisfy the requirement $|h_0\pm h_1|<1$.

The condition in Eq.~\eqref{4.4} can now be tested for validity.
With Eq.~\eqref{6.18}, the condition in Eq.~\eqref{4.4} becomes
\begin{equation}
\label{6.28}
|p (v_d^2-1) D_{d0}/4| = |p(v_d^2-1) / \gamma_d |<1 \, .
\end{equation}
Because $v_d^2 - 1$ and $\gamma_d$ have the same sign (positive for the triplet, negative for the singlet),
\begin{equation}
\label{6.29}
p < \gamma_d / (v_d^2 - 1) =
\begin{cases}
 181~\rm{MeV}/c & \ \rm{triplet} \\
 151~\rm{MeV}/c & \ \rm{singlet}
\end{cases} \ .
\end{equation}
These correspond to c.m. kinetic energies of 34~MeV and 24~MeV, well above the low-energy range for
which SI ERT is an excellent description of the data.
With the reasonable assumption that $h_0$ and $h_1$ (and perhaps even $D_{d0}$) are energy dependent,
and with
$v_d^2-1 = (h_0\pm h_1)(2-h_0\mp h_1) / (1-h_0\mp h_1)^2$,
it is possible that the condition in Eq.~\eqref{6.28} is always satisfied,
provided that $h_0, h_1 \to 0$, or $D_{d0} \to 0$, as $p \to \infty$.
\section{Conclusion and Discussion}
\label{sec7}
From Eqs.~\eqref{6.2}, \eqref{6.15}, and \eqref{6.25},
\begin{gather}
\label{7.1}
a_d^{-1} = \gamma_d(1-\tfrac{1}{2} \gamma_d r_d) \ , \quad
\left|\tfrac{1}{2} \gamma_d r_d \right| < 1 \ . 
\\ \label{7.2}
 \mbox{For} \ p=0: \ R_d = v_d^2 = (1-\tfrac{1}{2} \gamma_d r_d)^{-1} \ ,
\\ \label{7.3}
R_t = 1.252~90(35) \ , \ 
R_s = 0.9479(10) \ .
\end{gather}
The sign of $a_d$, which determines the sign of $\gamma_d$ through Eq.~\eqref{7.1} (or vice versa),
manifests itself in the shape of the spectrum through the term $\tfrac{1}{2} \gamma_d r_d$ in Eq.~\eqref{7.2}:
At zero energy $R_d>1$ for $a_d>0$ and $R_d<1$ for $a_d<0$.
But, by the mechanism proposed here, $R_d = v_d^2 q_d$ has precedence over $a_d$, rather than the other way around.
That is,
whether $R_d$ is greater or less than unity at zero energy determines the sign of
the scattering length.
From Eqs.~\eqref{3.4}, \eqref{7.2}, $1 \gg h_1 > h_0 > 0$, and $r_d > 0$,
\begin{equation}
\label{7.4}
\gamma_d =
\begin{cases}
 2 ( h_0 + h_1 ) ( 2 - h_0 - h_1 ) / r_d > 0 & \ \rm{triplet} \\
 2 ( h_0 - h_1 ) ( 2 - h_0 + h_1 ) / r_d < 0 & \ \rm{singlet}
\end{cases} \ .
\end{equation}
Thus, the sign of $\gamma_d$, 
{\em and therefore the sign of $a_d$},
is determined by pion exchange and the spatial symmetry of the free $np$ wavefunction:
The triplet amplitude is increased by $\pi^\pm$ exchange
because there is no sign change in the exchanged triplet $np$ wavefunction,
while
the singlet amplitude is reduced by $\pi^\pm$ exchange
because of the change in sign of the exchanged singlet $np$ wavefunction.
This behavior follows from Eq.~\eqref{3.4}, and is independent of the comparison with ERT in Sec. \ref{sec6}.\cite{attenuation}

Referring back to Fig. \ref{Fig2}, the first-order vertex correction, the $NN\pi$ transition elements are
$y_1=y_{nn'\pi^0}, \ y_2=y_{pp'\pi^0}$ for $\pi^0$ exchange, and
$y_1=y_{np'\pi^\pm}, \ y_2=y_{pn'\pi^\pm}$ for $\pi^\pm$ exchange.
A Coulomb correction $F_\pm$ is present in
$y_{np'\pi^-}$ ($F_+$) and $y_{pn'\pi^-}$ ($F_-$), where $p \pi^-$ occur together,
but not in
$y_{np'\pi^+}$ nor $y_{pn'\pi^+}$, where no charged particles occur together.
The $np'\pi^-$ and $pn'\pi^-$ vertices always appear together, so only the product $F_+ F_-$ occurs.
Nonrelativistically \cite{134},
\begin{gather}
\label{7.5}
F_+ F_- = \pi^2 \eta^2 / \sinh^2 \pi \eta \ ,
\end{gather}
where
$\eta \equiv \alpha / \beta$ is the Sommerfeld parameter, $\beta$ is the $p\pi^-$ relative velocity,
and $\alpha\cong 1/137$ is the electromagnetic coupling.
$F_+ F_-$ is very nearly unity, except for low pion-proton relative velocities,
where $F_+ F_- \to 0$.
The effect is to attenuate the contribution from $\pi^-$ exchange compared to
the contributions from $\pi^0$ and $\pi^+$ exchanges.
There being two charged and one neutral pion,
ignoring differences in the pion-nucleon couplings
we would have $h_1=2h_0$.
However, because the coulomb correction attenuates the contribution from the $\pi^-$,
$h_1$ should be somewhat smaller than twice $h_0$.
Let $C_\pi \lesssim 1$ be a factor accounting for the Coulomb correction.
Then
\begin{equation}
\label{7.6}
h_1 = h_+ + h_- = h_0+C_\pi h_0 \ ,
\end{equation}
where the $h_+=h_0$ term applies to $\pi^+$ exchange (no Coulomb correction)
and $h_- = C_\pi h_0$ applies to $\pi^-$ exchange (Coulomb correction).
Then
\begin{equation}
\label{7.7}
C_\pi  = (h_1/h_0) - 1 = \, 0.682(14) \ .
\end{equation}
Any differences (besides the Coulomb correction) in the pion-nucleon coupling strengths
of $pp\pi^0$, $nn\pi^0$, and $np\pi^\pm$, such that
$h_+ \neq h_0$ for example, are unavoidably included in $C_\pi$,
but that should be a small part of the value shown in Eq.~\eqref{7.7} for low energies.

From Eqs.~\eqref{6.15} and \eqref{7.2}, (upper sign for the triplet)
\begin{align}
\label{7.8}
a_d &= v_d^2 / \gamma_d = 1 / \gamma_d (1 - h_0 \mp h_1)^2 \ ,
\\ \label{7.9}
r_d &= 2 \left(\! 1- v_d^{-2} \right) / \gamma_d
= 2 (h_0 \pm h_1) ( 2 - h_0 \mp h_1 ) / \gamma_d \ .
\end{align}
The extension from the shape-independent ERT to the shape-dependent ordinarily involves obtaining
the coefficients of $p^4$ and higher orders of $p^2$ in the expression for the phase shift
$p \cot \delta_d = -1/a_d + \tfrac{1}{2} r_d \, p^2 - P_d \, p^4 + ...$, but if the {\em exact} expression for the
effective range $r_d$ is employed \cite{366,2000}, then the higher-order terms are accounted for in the
energy dependence of $r_d$.
From Eq.~\eqref{7.9}, the condition for shape-independence, or constant $r_d$, is that $h_0$ and $h_1$
have no energy dependence, i.e., the energy domain over which the SI ERT is an excellent description
is coincident with the domain over which
the energy dependence of the nucleon-pion interaction is negligible.

Presumably, $h_0$ and $h_1$ are energy-dependent,
which should become noticeable somewhere above 3-10~MeV with the currently available data.
Therefore, $a_d$ and $r_d$, as given by Eqs.~\eqref{7.8} and \eqref{7.9},
also depend on the energy, through $h_0$ and $h_1$.
While $r_d$ depends on $h_0\pm h_1$ at the lowest order,
$a_d$ depends on $h_0 \pm h_1$ only at the next order:
At higher energies, $a_d$ is much better approximated by a constant than is $r_d$.
Note that Eqs.~\eqref{7.8} and \eqref{7.9} are not truncated expansions;
they permit an extension to higher energies,
if only the energy dependence of $h_0$ and $h_1$ were determined,
either from data or through {\em ab initio\/} calculations of $h_0$ and $h_1 = h_- + h_+$,
including non-pionic contributions.
For exchange of mesons with spin, where the meson-nucleon coupling is {\em not\/} spin-independent,
$h_0$ and $h_1$ will acquire contributions which differ for the triplet and the singlet.
At higher energies, the $d$-wave in the triplet propagator and its correction would also need to be treated,
as well as the appearance of heavier baryons in the propagator and vertex corrections.

Because $v_t$ and $v_s$ are well-approximated by constants at low energy,
the infinite-series argument leading to Eq.~\eqref{3.3}, i.e., $[1-h_0\mp h_1]^{-1}$,
could have been replaced with a one-pion-exchange approximation, yielding instead $1+h_0\pm h_1$,
and slightly different values for $h_0$ and $h_1$.
The conclusions drawn herein regarding the scattering length sign and spatial symmetry of the free $np$ wavefunction
remain the same in either case, because $h_1 > h_0$.
However, with a one-pion-exchange approximation, the derivation leading to Eq.~\eqref{4.1}
would not have been so clear-cut.

Quite the contrary with Eq.~\eqref{4.3}, because of its explicit energy-dependence.
Only the form in Eq.~\eqref{4.3} produces a result identical to that of the SI ERT.
Simply taking the leading term in the series yields, instead of Eq.~\eqref{4.5}, $q_d = 1 + i p(v_d^2-1)D_{d0}/4$,
which performs poorly at low energies, whereas Eq.~\eqref{4.5} continues to perform fairly well even at energies
that violate Eq.~\eqref{6.29}.
No obvious line of reasoning produces the form Eq.~\eqref{4.3}, other than the infinite series one.
No simple energy-dependent forms for $h_0$, $h_1$, and $D_{d0}$
seem to satisfy the condition in Eq.~\eqref{6.29} for higher energies, without also compromising the low-energy
agreement with ERT.
In particular, taking $D_{d0} = \tfrac{1}{2}D_d$ guarantees that Eq.~\eqref{6.29} is always satisfied,
but performs poorly at low energies.
Likewise, simple forms such as $h_0(p) = h_0(0) / [1+p h'_0]$, do not have sufficient energy-dependence
to rescue Eq.~\eqref{6.29}.
Threshold effects, which are certainly present,
would rule out any simple energy-dependent forms for $h_0$, $h_1$, and $D_{d0}$.
(The same argument can be applied to the standard shape-dependent expansion --
threshold effects should rule out simple, constant values for the coefficients $P_d$, etc.)
In any case, because Eq.~\eqref{4.3} works so well at low energies
and because it does not fail catastrophically at higher energies
as it should if Eq.~\eqref{6.29} were really required,
it may be that the condition in Eq.~\eqref{6.29} applies to the {\em validity of the derivation}, but not to Eq.~\eqref{4.3} itself.

%
\section{Extension:  $pp$ and $nn$ scattering}
\label{sec8}
To fully treat $pp$ scattering by the diagrammatic approach presented here would
require the introduction of the Coulomb barrier (e.g. \cite{199,1348})
\begin{equation}
\label{8.1}
C \equiv 2\pi\eta / \left( e^{2\pi\eta} - 1 \right) \, ,
\qquad
\eta \equiv \alpha/\beta \, ,
\end{equation}
where $\beta$ is the relative $pp$ velocity.
The problem is further complicated by interference between the Coulomb and nuclear interactions.
However, because there are available in the literature values of the ERT $pp$ scattering parameters $a_{pp}$ and $r_{pp}$
which have had the Coulomb effect removed,
these should be comparable to ones that could be obtained with the method used here,
without introducing the Coulomb barrier and its attendant complicating interference effects.
On the other hand, the $nn$ scattering parameters should follow directly from the results presented here for $np$ scattering,
notwithstanding the fact that the $nn$ parameters are somewhat indirectly obtained,
owing to the impracticality of directly performing $nn$ scattering experiments.
For both $pp$ and $nn$ scattering, only the spin-singlet can contribute.

For $np$ scattering, $\pi^0$ exchange does not result in a change of identities between the neutron and proton;
that is, the $n$ and $p$ do not change places, hence Eq.~\eqref{3.1}.
For $nn$ or $pp$ scattering, however, $\pi^0$ exchange {\em does\/} produce such a change
because the particles are indistinguishable fermions,
and the spin-singlet wavefunction therefore changes sign.
Instead of Eq.~\eqref{3.1}, one expects
\begin{equation}
\label{8.2}
H_0 |pp\big> = -h_0|pp\big> \, , \qquad
H_0 |nn\big> = -h_0|nn\big> \, .
\end{equation}
Charged $\pi$ exchange is not possible, so there is no equivalent to Eq.~\eqref{3.2}.
If the $np\pi^0$ coupling were the same
as that of $pp\pi^0$ and $nn\pi^0$ (i.e., charge independence and charge symmetry),
then $h_0$ would be the same for $np$, $nn$, and $pp$.
The vertex correction for both $nn$ and $pp$ would then be given by Eq.~\eqref{3.4},
if $h_0$ occurred with opposite sign and if $h_1$ did not occur at all, thus
\begin{equation}
\label{8.3}
v_{pp} = v_{nn} = (1 + h_0)^{-1} = 0.961~77(26) \, .
\end{equation}
Because $v_{pp} = v_{nn}<1$, the $pp$ and $nn$ scattering lengths should be negative, as they are observed to be.
Instead of Eqs.~\eqref{7.8} and \eqref{7.9}, we have
\begin{gather}
\label{8.4}
a = 1/\gamma(1+h_0)^2 \, , \quad r = -2h_0(2+h_0)/\gamma \, .
\end{gather}
The $nn$ and $pp$ ERT parameters (with Coulomb and magnetic effects removed) from \cite{1314} are
\begin{align}
\nonumber
& a_{pp} = -17.3(4)~\mbox{fm}
\quad
& r_{pp} = 2.85(4)~\mbox{fm} \, \phantom{1.}
\\
\label{8.5}
& a_{nn} = -18.8(3)~\mbox{fm}
\quad
& r_{nn} = 2.75(11)~\mbox{fm} \, .
\end{align}
The model presented here makes no prediction of the scattering pole masses,
or equivalently $\gamma_{pp}$ and $\gamma_{nn}$.
Without fitting to data (which would greatly exceed the scope of this work),
there is no choice but to obtain those from the ERT parameters above.
These give
\begin{equation}
\nonumber
\gamma_{pp} = -0.0537(12)~\mbox{fm}^{-1}
\quad
\gamma_{nn} = -0.04978(77)~\mbox{fm}^{-1} \, .
\end{equation}
\smallskip
With those, Eq.~\eqref{8.4}, and $h_0$ from Eq.~\eqref{6.26},
\begin{align}
\nonumber
& a_{pp} = -17.23(38)~\mbox{fm}
\quad
& r_{pp} = 3.020(70)~\mbox{fm} \, \phantom{.}
\\
\label{8.6}
& a_{nn} = -18.58(29)~\mbox{fm}
\quad
& r_{nn} = 3.257(56)~\mbox{fm} \, .
\end{align}
While the agreement between these scattering lengths and those in Eq.~\eqref{8.5} is quite good,
the disagreement between the effective ranges is substantial.
It is well-known that charge symmetry breaking (CSB) and charge independence breaking (CIB)
have a larger effect on the effective ranges than on the scattering lengths.
This is precisely what is happening here, and it is because the effective range depends on the vertex corrections
at the lowest order, whereas the scattering lengths depend on them only at the next order, as mentioned in Sec. \ref{sec7}.

In terms of the scattering length and effective range, the vertex corrections for $pp$ and $nn$ are given by
\begin{equation}
\label{8.7}
v = \sqrt{ (a/r) ( 1 - \sqrt{1-2r/a} ) } \, .
\end{equation}
With $h = (1/v) - 1$ and the parameters in Eq.~\eqref{8.5} from \cite{1314},
\begin{equation}
\label{8.8}
h_{pp} = 0.03755(58) \qquad h_{nn} = 0.03366(41) \, .
\end{equation}
\vskip 0 pt
\noindent
The difference between these and $h_0$ from $np$ in Eq.~\eqref{6.26} is 3.4 times the $np$ and $pp$ combined errors for $pp$,
12 times the $np$ and $nn$ combined errors for $nn$,
and the difference between $h_{pp}$ and $h_{nn}$ is 5.5 times their combined errors.
This is to be compared with the much smaller relative difference (dividing by the corresponding combined error)
between any combination of $r_s$ for $np$ from Eq.~\eqref{6.24}
and $r_{pp}$ and $r_{nn}$ from Eqs.~\eqref{8.5} and \eqref{8.6}.
The isolated $\pi^0$ exchange eigenvalues $h_0$, $h_{pp}$, and $h_{nn}$, then,
are even more sensitive to CSB and CIB than are the effective ranges.

%
\begin{acknowledgments}
I would like to thank
R. Chrien,
S. U. Chung,
and
L. Trueman
of BNL,
and G. Hale of Los Alamos,
for informative discussions.
I am grateful to S. Borzakov of JINR, Dubna, Russia, for drawing
my attention to his and other works.
I would also like to thank L. Littenberg for his assistance.
This work was supported by the United States Department of Energy, under Contract No. DE-AC02-98CH10886.
\end{acknowledgments}

%
%
%

%

\end{document}